\documentclass[12pt]{iopart}
\usepackage{graphicx}

\begin{document}

\title[Electrodiffusion in ice]{An electrodiffusion model for Jaccard's theory in ice}

\author{John De Poorter}

\address{Horatio vzw, Ghent, Belgium} 
\ead{john.zarat@gmail.com}
\vspace{10pt}
\begin{indented}
\item[]June 2020
\end{indented}

\begin{abstract}
Jaccards' theory describes the movement of both ionic and Bjerrum defects in ice. Standard descriptions of the theory are based on a chain model describing the movement of these defects along well-oriented chains of water molecules. However, this model contains several fundamental contradictions and does not result in the exact equations. We present an alternative model based on the electrodiffusion of the defects. The polarisation of the ice specimen favours these defects orientations that diffuse opposite to the electric drift of the same defect. This straightforward approach not only results in the correct equations, it also provides a better understanding of the defects' kinetics. 
\end{abstract}

\vspace{2pc}
\noindent{\it Keywords}: Jaccards' theory, ice, electrodiffusion

\submitto{\EJP}
%\maketitle

\section{Introduction}
\label{sec:intro}

Electrodiffusion occurs when charged particles move under the combined influence of both an electric field and diffusion~\cite{RN1111}. However, in the standard ice-physics literature ~\cite{RN10943, RN13131,RN10877,RN1107}, the movements of the defects in ice (both ionic and Bjerrum defects) are not linked to a diffusion process. Instead, a chain model is used based on the opening and closing of chains of water molecules by the different types of defects. This model didn't allow Jaccard to find the correct flux equations of the defects in ice~\cite{RN10943} and we will show in this paper that it contains several fundamental contradictions. 

The original theory of Jaccard was already corrected in \cite{jdp1} because it made no distinction between bound and free charges. The lack of this distinction leads to some didactical problems in the theory like the ad hoc introduction of the so-called configuration vector~$\Omega$. When the two types of charges are distinguished, it becomes clear that $\Omega$ is redundant and proportional to the polarisation density. The analytical equations for the defect flux densities were calculated as the sum of an electric and a diffusion component. So, in this paper a first version of an electrodiffusion model for ice can be found. However, the diffusion contribution was derived as a concentration difference between the defects. Although correct formula were found, a constant and significant macroscopic concentration gradient of the defects should be present throughout the ice slab (the concentration should double every 5 mm for an electric field of 10 V/m). Such a strong gradient is very unlikely and is not observed experimentally. We therefore reworked this paper and found that the diffusion gradient is not induced by a concentration gradient but by the building up of a preference orientation of the defects due to polarisation of ice. Defects with this orientation are blocked in their Brownian movement in the direction of the electrical drift of the same defects, creating a netto diffusive current opposite to the drift current. We will also extend the ice equations for gradients in the defect concentration and the polarisation density. The extension is valuable for the study of both boundary effects~\cite{RN1107, RN11870} and thermoelectrics in ice~\cite{RN1110,RN1109}. 

\section{Components of the model}

\subsection{A 2D model for ice}

\begin{figure}
\centering{\includegraphics[width=100mm]{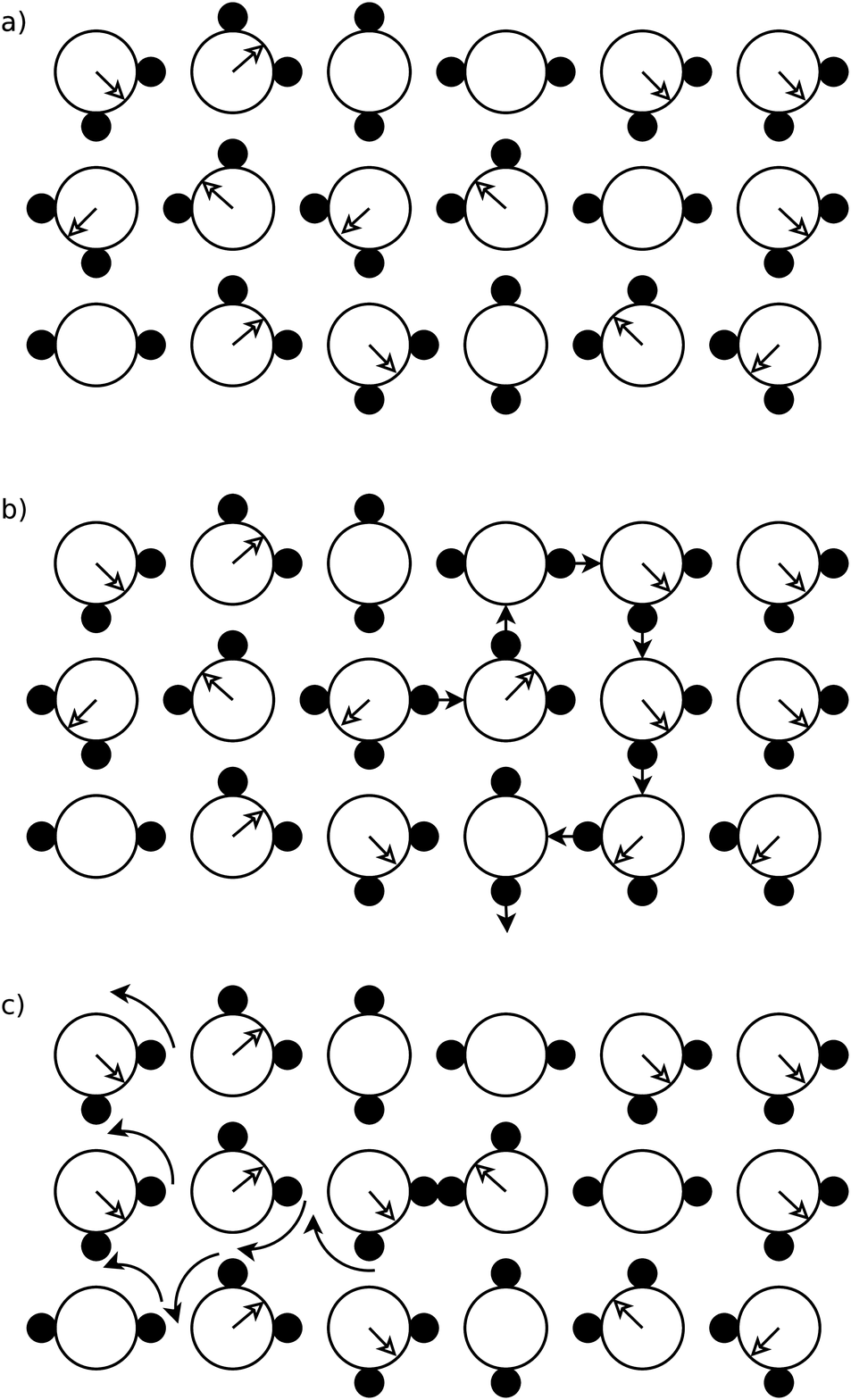}}
\caption{ (a) 2D representation of the ice lattice. The white circles are oxygen atoms, the black ones hydrogen atoms. The dipole moment of the individual water molecules $\mathbf{p_o}$ is indicated by the small arrows inside the oxygen atoms. (b) A H$^+$ defect is jumping randomly throughout the lattice. (c) The random movement of a Bjerrum D defect.  }
\label{fig:2Dmodel1}
\end{figure}
 
The most common crystalline phase of ice, ice I$_h$, is hexagonal. Each water molecule is fixed in the ice crystal structure and surrounded by 4 nearest neighbours located at the corners of a tetrahedron~\cite{RN10913}. There are two rules describing the orientation of individual water molecules in the crystal structure, the so-called Bernal-Fowler ice rules. The first rule is that each molecule accepts two hydrogen atoms from two nearest-neighbours water molecules and also donates two hydrogen atoms to the two other nearest-neighbours. The second rule states that there is precisely one hydrogen atom between each pair of oxygen atoms. As an illustration of these rules, we have drawn a 2D version of a defect-free ice lattice in Fig.~\ref{fig:2Dmodel1}(a)~\cite{jdp1}.

The horizontal x axis is chosen as the potential direction for an external field. The dipole moments of the water molecules, projected on this axis, have a size called $p_{o,\parallel}$ and they can be positive or negative depending on the direction of the water molecules related to the x axis. We define $n_{p^+}$ and $n_{p^-}$  as the density of water molecules polarised in the positive and negative x direction, respectively and $n_{p=o}$ as density of the water molecules with a zero polarisation component in the x direction.  Notice that 
\begin{equation}
 n_o = n_{p^+} + n_{p^-}+n_{p=o} \label{eq:noo} 
\end{equation}  
and that in the absence of an external electric field $n_{p^+}=n_{p^-}=n_{p=o}=n_o/3$ resulting in a zero polarisation of the specimen. If an external field is applied in the direction of the positive x axis, the polarisation density of the ice lattice $\mathbf{P_s}$ in the horizontal axis is by definition equal to
\begin{equation}
\mathbf{P_s}=  (n_{p^+}-n_{p^-})p_{o,\parallel}  \mathbf{e_x},   \label{eq:Pl} 
\end{equation}  
with $\mathbf{e_x}$ the unit vector in the x direction. There are two kinds of contributions to the polarisation density. The first contribution is related to the displacement of the charge distribution of individual water molecules under influence of an external field and is called the water molecule polarisation $\mathbf{ P_m}$. $\mathbf{ P_m}$ is only a small fraction (only some procent) of $\mathbf{ P}$, the total polarisation density of ice~\cite{RN10943}. The lattice polarisation $\mathbf{ P_s}$ provides the major part of the polarisation. It is caused by the netto orientation of the permanent dipole moments $p_o$ of the water molecules. We focus only on this contribution. 

\subsection{Four types of defects}
When an electric field $E$ is applied to ice, the ice lattice becomes easily polarised and has a large relative dielectric constant (93 at {-}$3^\circ$C)\cite{RN8152}. This effect is due to the reorientation of a fraction of the individual watermolecule dipoles in the direction of the applied field. However, this reorientation can only happen if the ice rules are at least temporary broken. Therefore, mobile lattice defects are assumed to be present inside the ice crystal changing the orientations of the water molecules when they move. 

Two types of defects are described in ice: ionic defect pairs which violate the first ice rule and the Bjerrum defect pairs embodying the violation of the second ice rule~\cite{RN10943}. A H$^+$-OH$^-$ ionic defect pair is created when one of the hydrogen atoms jumps to the neighbouring water molecule leaving its electron behind. Both ions can separate and move independently throughout the lattice. In Fig.~\ref{fig:2Dmodel1}(b) the movement of a H$^+$ ion is visualised. The second ice rule is violated by turning one water molecule (or a hydrogen atom of the molecule) over an angle (90$^\circ$ in a 2D lattice) so that one of its neighbouring O-O bonds is occupied by two hydrogen atoms (a Bjerrum D defect) and the other one with no hydrogen atoms (a Bjerrum L defect). The Bjerrum defects can separate and move independently throughout the lattice, a moving D defect is visualised in Fig.~\ref{fig:2Dmodel1}(c). The DL defects are seen as quasi particles because they behave in a way similar to a real charged particle \cite{RN11906}. However, they are not real physical particles but only a temporary deviations of the Bernal-Fowler rule in the ice structure. 

The ice specimen is examined using the macrocopic Maxwel equations, thereby replacing matter by bound charges induced by spatial changes in the polarisation density  $\mathbf{ P_s}$~\cite{RN13018}. The density of bound charges $\rho_b$ can be found using
\begin{equation}
\rho_b = - \vec{\nabla .} \vec{P_s} \label{eq:Maxwell1}
\end{equation} 
Because the dipole moments change their orientation over both Bjerrum defects and ionic defects (see Fig.~\ref{fig:2Dmodel1}), the netto charge of both type of defects can be described using bound charges. It is shown in~\cite{jdp1} that a D defect has a bound charge $e_{DL}$ ($= 0.38 e$) and an L defect $-e_{DL}$ ($= -0.38 e$). The ions contain both the free charge $\pm e$  and an opposite bound charge $\mp e_{DL}$, resulting in a netto charge named $\pm e_\pm$ ($= \pm 0.62 e$) for the positive and negative ions respectively. Table~\ref{Table1} summarises the charges of the four types of defects. 

\begin{table}
\caption{ The free and bound charges of the four types of defects. Also, their contribution to the bound current density ($\mathbf{J_b}$) and the free  current density ($\mathbf{J_f}$) are tabulated. $\mathbf{j_+}$, $\mathbf{j_-}$, $\mathbf{j_D}$, and $\mathbf{j_L}$ are the defect flux densities with the index referring to the type of defect. }
\label{Table1}
\begin{tabular}{   c  || c   c    c   c   c}
\hline 
defect& free &bound &total  &$\mathbf{J_b}$ &$\mathbf{J_f}$\\
\hline 
H$^+$ & e& $-e_{DL}$&$e_{\pm}$ & $-e_{DL} \mathbf{j_{+}}$ &$e \mathbf{j_{+}}$\\
OH$^-$ & $-e$& $e_{DL}$& $-e_{\pm}$ & $e_{DL} \mathbf{j_{-}}$ &$-e \mathbf{j_{-}}$\\
D &  0& $e_{DL}$& $e_{DL}$ &  $e_{DL} \mathbf{j_{D}}$ &0\\
L &  0& $-e_{DL}$&$-e_{DL}$ &  $-e_{DL}\mathbf{ j_{L}}$ &0 \\
\hline 
\end{tabular}
\par
\end{table}

\subsection{The misleading chain model for defect motion}

The electric conductivity in ice has uncommon properties. At first sight both types of defects should contribute to it, but experimental data shows that the DC conductivity is only determined by the type of defects with the lowest concentration~\cite{RN10943}. Increasing the concentration of the most abundant type of defect (the ions or the DL's) will not alter the conductivity of the ice significantly. This property is commonly explained using a chain model for the defects movement. 

The ionic movement is based on proton jumping from one water molecule to the other, the so-called Grotthus mechanism~\cite{RN11914}. In the 2D lattice the ions move in both the horizontal $x$ direction and the $y$ direction. In the absence of an external electric field the ionic movement will be random in all directions. A typical movement for the H$^+$ is visualised in Fig.~\ref{fig:2Dmodel1}(b). 

\begin{figure}
\centering{\includegraphics[width=130mm]{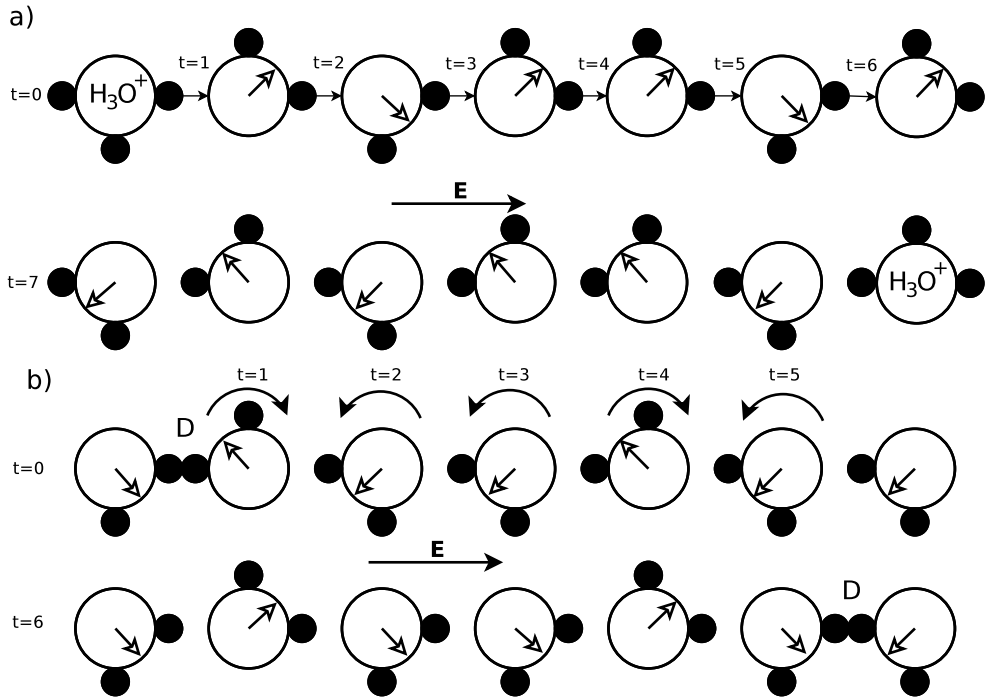}}
\caption{2D representation of the movement of defects in ice through well-oriented chains under the influence of an external electric field $\bf{E}$. First row: a H$^+$ defect (a) or a D defect (b)  is present at the left side of a well-chosen chain of water molecules at t = 0. Second row: the orientation of the water molecules after the movement.} 
\label{fig:2Dmodel2} 
\end{figure}

In the chain model, the H$^+$ ion will move under the influence of an external electric field through a well-oriented chain in the water structure~\cite{RN11914}. The 2D versions of this model is visualised in Fig.~\ref{fig:2Dmodel2}(a) (a 3D version of these chains can be found in the work of Jaccard~\cite{RN13131}). In the upper row of (a), the H$^+$ ion jumps through a chain of water molecules with dipole moment oriented in the direction of the electric field. Notice that these consecutive jumps result in the lower chain consisting of water molecules polarised opposite to the electric field. As a consequence, the passing of one H$^+$ prevents other H$^+$ ions to move in the same path. 
The electric susceptibility of a theoretical ice slab with significantly more ionic defects than DL defects should be negative. However, both natural and prepared ice slabs have more DL defects and a positive susceptibility~\cite{jdp1}. 

A similar reasoning can be made for the DL defects. A random movement for the D defect in the absence of external field is visualised in Fig.~\ref{fig:2Dmodel1}(c). The chain view on the movement under influence of an external field is presented in Fig.~\ref{fig:2Dmodel2}b. In the  upper row, the water molecules are oriented in the direction opposite to the electric field. The driving force for this movement is the netto torque on the water molecules trying to orient themselves with a dipole moment parallel to the electric field. So the passing by of a D defect from left to right, will also change the orientation of the water molecules and also here, this path is blocked for other D defects. 

Jaccard describes the conduction mechanism by the closing and opening of chains of water molecules through which defects can propagate~\cite{RN13131}. The central idea is that the propagation of one type of defect in the lattice closes this chain for other defects of the same type. At a first glance, this explains why one type of defect cannot propagate throughout ice: it is closing the structure for the other defects. The chains can only be reopened by another type of defects explaining why the DC current through an ice slab is determined by the defects with the lowest concentration. The excess of the other defects can't move because their chains are closed. This model is both visual and intuitive and became the standard explanation~\cite{RN10943}. However, it offers a misleading view on what is actually happening. 

First of all, the Bernal-Fowler ice rules exclude the possibility that there are two similar defects (e.g. two H$^+$ or two D defects) in one chain. So the reasoning that the movement of one defect is closing the structure for other similar defects is not relevant at all. Second, the number of defects is assumed to be small making it improbable that a H$^+$ (or OH$^-$) defect and a D (or L) defect are on the same chain. Thirdly, the chain model neglects the 3 dimensional character of the defect movement. If a chain is blocked, the defects will move in a direction perpendicular of the applied field (like in Fig.~\ref{fig:2Dmodel1}(b) en (c)) to another open chain. The argument that after some time all the open chains are blocked is also no way out. The polarisation of dielectrics results in only a very limited extra fraction of the water molecules that become oriented in the direction of the external field (at -10$^\circ$C an electric field of 1000 V/m will only flip 3 ppm of the water molecules in ice). An even more fundamental fourth argument points out that the whole concept of a long chain doesn't reflect the entropy of a real ice crystal. In the 2D lattice this is reflected by the fact that one third of the water molecules have a zero polarisation component in the x direction and will be limiting the length of the chains. If we add the fact that Jaccard didn't succeed to use his kinematic model based on the chain model to derive the correct equations (he used general thermodynamical considerations instead)~\cite{RN10943}, we may conclude that the chain picture is blocking a deep understanding of the real physics inside the ice crystal.

\subsection{An electrodiffusion model for defect motion}
\label{electrodiffusion}
The main idea of this new model is that flux density enforced by the applied electric field is counterbalanced by a diffusive flux of the same defects. If the lattice has no external field and no netto orientation of the dipole moments ($P_s = 0$), there is no netto diffusive displacement of defects. In this section we will show that the ohmic currents will create a dominant orientation in the water molecules ($P_s  \neq 0$) inducing a preferable direction in the thermal hopping of the defects, opposite to the electric drift. Within this view, the electric and the diffusion currents will compensate each other completely without the necessity of the closing of specific chains. So one type of defects cannot contribute to the conductivity alone, explaining the experimental findings. 

\begin{figure}
\centering{\includegraphics[width=90mm]{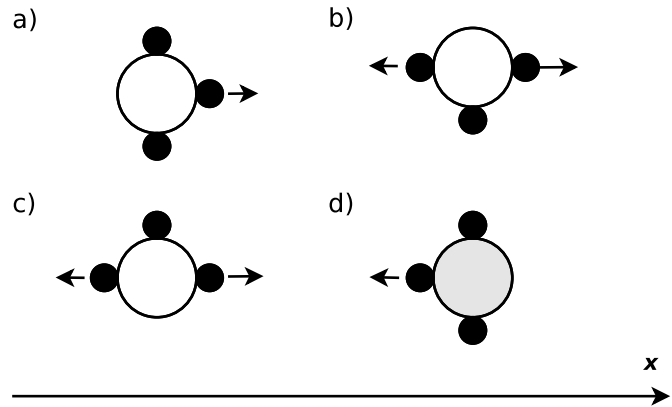}}
\caption{2D representation of the different orientations (a-d) of the  the H$^+$ ions and how the complete ions can move in the x direction. The grey H$^+$ ion is blocked in the positive x direction (case d), the direction of the external electric field.} 
\label{fig:2Dmodel5} 
\end{figure}

The details of the diffusion mechanism are first illustrated for a specimen with only ionic defects. The way a H$^+$ defect is hopping throughout an ice structure without an external electric field is already visualised in Fig.~\ref{fig:2Dmodel1}(b). Fig.~\ref{fig:2Dmodel5} illustrates how the orientation of the  H$^+$ ion is limiting the possible movements of the ion in the $x$ direction. Notice that each H$^+$  ion can only move in three of the four possible directions. In case (d) the movement of the H$^+$ ion is blocked in the positive x direction, which is the direction the external electric field will be applied. However, in an unpolarised ice specimen ($n_{p^+}=n_{p^-}=n_{p=o}=n_o/3$), the four orientations are equally present. So, no effect of this asymmetry is found. The hopping will be equal in all the directions, there is no netto current. 

If an electric field is applied in the positive $x$ direction, there will be an electric current in the same direction. The ion shown in Fig.~\ref{fig:2Dmodel5}(d) may induce a diffusional current in the negative x direction. This will not happen initially, because ions of orientation (a) compensate for the movement of ions of orientation (d). However, the ice specimen becomes more and more polarised by the electric current. When only ions are present, this polarisation will be opposite to the electric field (see Fig.~\ref{fig:2Dmodel2}(a), $n_{p^+}<n_{p^-}$) and water molecules with a dipole moment opposite to the electric field cannot form (a) type ions.  So, the orientation (d) will occur more frequently than the (a) orientation resulting in a netto diffusional ionic current opposite to the electric ohmic current of the ions. This diffusive current will grow until both currents compensate each other. 

\begin{figure}
\centering{\includegraphics[width=110mm]{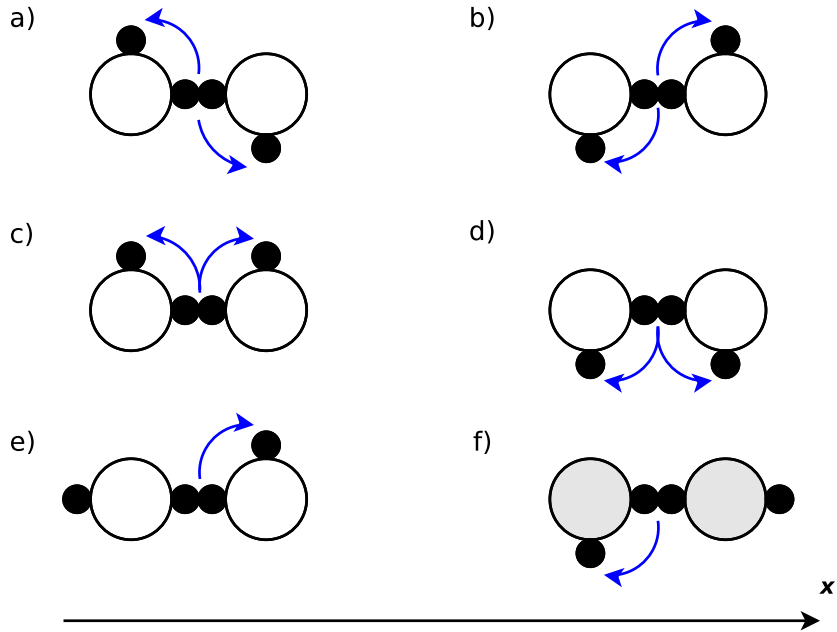}}
\caption{2D representation of the different orientations (a-f) of horizontal D defects. We only indicated the rotations allowing the D defect to move in the x direction. Only the grey D defect is blocked in the positive x direction (case f), the direction of the external electric field. This type of D defect will be more present when the ice specimen is polarised by the electric D current.} 
\label{fig:2Dmodel6} 
\end{figure}

A similar reasoning can be made for an ice specimen with only Bjerrum defects. A D defect is composed of two water molecules and depending on the specific orientation of these molecules 6 different types of horizontal D defects can be found (see Fig.~\ref{fig:2Dmodel6}). In cases (a) to (d) the D defects can move in both directions. In case (f) the D defect is blocked in positive $x$ directions. These orientations are more favoured in an ice specimen that is polarised in the same direction as the electric field (see Fig.~\ref{fig:2Dmodel2}(b), $n_{p^+}>n_{p^-}$). The D defects will therefore diffuse in the direction opposite to the electric field thereby compensating the electrical current at a certain level of polarisation.

A last question remains unanswered. How can a current of two different types of defects be stable if both defects individually cannot build up a stable electric current? A netto positive current of the ions will polarise the ice specimen opposite to the electric field, while a netto positive current of the DL defects will compensate this opposite polarisation. So a balance between both type of currents will prevent extra polarisation of the ice specimen. This balance allows a stable DC current and a DC conductivity mainly determined by the type of defects of the lowest concentration.

\subsection{Derivation of the ice-flux equations}
In this section, the electrodiffusion of the defects will be quantified resulting in the well-known ice equations. Our approach will allow us to extend the classical equations to the case of nonuniform electric fields and defect concentrations. 

We will first focus on the detailed derivation for the H$^+$ ions. 
$\sigma_+$ is the conductivity of the positive H$^+$ ions in an ice specimen,  
\begin{equation}
\sigma_+ = e_\pm n_+ \mu_+ . \label{eq:s1}
\end{equation} 
with $n_+$ the particle density, $\mu_+$ the H$^+$ mobility and $e_\pm$ the charge transported by the ion. The flux density of the H$^+$ ions (i.e. the number of ions crossing a unit area per unit of time) is denoted by the vector $\bf{j_{+}}$. This flux density differs from the current density (denoted with a capital letter $\bf{J_{+}}$) because it quantifies the number of ions passing by while the current density quantifies the net charge. When an electric field is applied in the horizontal direction, the H$^+$ flux density in the horizontal x direction is described by
\begin{equation}
j_+ = \frac{\sigma_+  E }{e_{\pm}}  + j_{+}^\leftrightarrow . \label{eq:fluxD1} 
\end{equation}
The first term is an ohmic drift term, the second is the diffusion term ($j_{+}^\leftrightarrow$) induced by the polarisation of the ice specimen and introduced in the previous section.

\begin{figure}
\centering{\includegraphics[width=110mm]{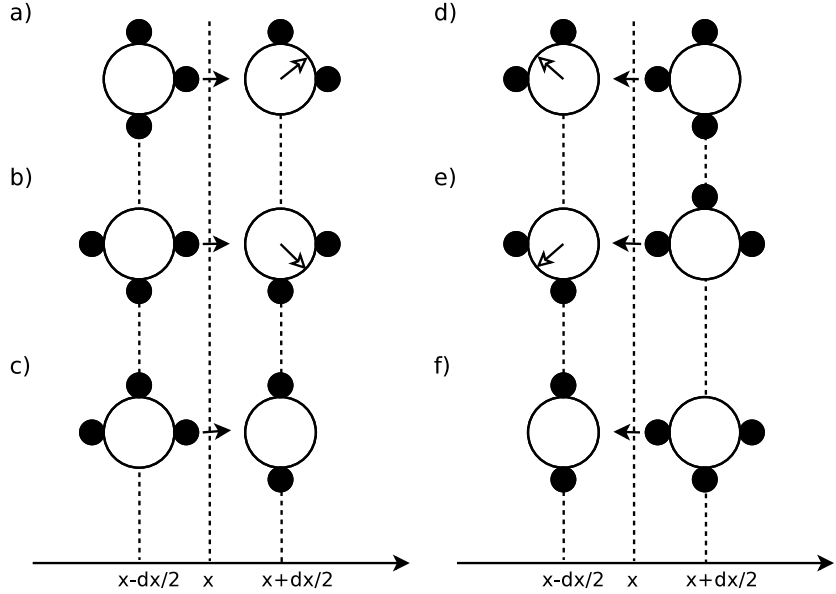}}
\caption{This 2D model shows the different ways a H$^+$ ion can move in the positive x direction (a-c) and in the negative x direction (d-f).}
\label{fig:2Dmodel3} 
\end{figure}    

We locate position $x$ just in between two water ions (see Fig.~\ref{fig:2Dmodel3}), $dx$ is the distance between the different planes of water molecules. The diffusional flux density of the H$^+$ defects at position $x$ is composed of two components, one from left to right $j_{+}^{\rightarrow} $ and one in the opposite direction from right to left $j_{+}^{\leftarrow} $.  $j_{+}^{\rightarrow} $ is initiated by the ions on position $x-dx/2$ (see Fig.~\ref{fig:2Dmodel3}a-c). In the previous section we focused on the blocked orientations in order to get a qualitative understanding of the two currents. In this quantitative approach, we count the unblocked H$^+$ ions. These ions can only jump to position $x+dx/2$ when the water molecules at this position have a dipole moment in the positive x direction (case a and b) or a zero dipole moment with both hydrogen atoms in the y direction (case c). So only the fraction 
\begin{equation}
 \frac{n_{p^+}(x+\frac{dx}{2}) +\frac{n_{p=o}(x+\frac{dx}{2})}{2}}{n_o}
\end{equation}
of the ions at position $x-dx/2$ has the correct orientation to allow a jump in the positive x direction. This leads to the following expression for $j_{+,D}^{\rightarrow} $ at position x
\begin{equation}
j_{+}^{\rightarrow}(x)  = n_+(x-\frac{dx}{2})  \frac{n_{p^+}(x+\frac{dx}{2}) +\frac{n_{p=o}(x+\frac{dx}{2})}{2}}{n_o} v_+(x),\label{eq:jr}
\end{equation}
with $v_+$ the average speed of the moving ions.

A similar reasoning can be held for the flux density in the negative x direction (see Fig.~\ref{fig:2Dmodel3}d-f). 
\begin{equation}
j_{+}^{\leftarrow}(x)  = n_+(x+\frac{dx}{2})  \frac{n_{p^-}(x-\frac{dx}{2}) +\frac{n_{p=o}(x+\frac{dx}{2})}{2}}{n_o} v_+(x). \label{eq:jl}
\end{equation}   

In order to simplify Eqs.~\ref{eq:jr} and \ref{eq:jl}, first order approximations for $n_+$, $n_{p^-}$,  $n_{p^+}$ and $n_{p=0}$ at positions $x\pm dx/2$ are used, e.g.  
\begin{equation}
n_+(x\pm \frac{dx}{2}) = n_+(x)\pm \frac{dx}{2}  \frac{\partial n_{+}}{\partial x}(x).
\end{equation}   
Notice that within the model of Fig.~\ref{fig:2Dmodel3}, $n_+(x)$ is not a physical quantity but a mathematical one easily derived assuming spatial continuity, 
 \begin{equation}
n_+(x) = \frac{n_+(x- \frac{dx}{2}) + n_+(x+ \frac{dx}{2})}{2}. 
\end{equation}  
This definition is consistent with the first-order approximations. 
  
Applying the first-order approximations and ignoring second order terms in $dx$, the netto diffusion current $j_{+, D}$ can be calculated with all quantities at that same position x,  
\begin{eqnarray}
j_{+} &= & j_{+, D}^{\rightarrow} - j_{+, D}^{\leftarrow}  \nonumber \\
&= &  v_+  \frac{n_+}{n_o} (n_{p^+}-n_{p^-}) -  v_+ \frac{dx}{2} \frac{\partial n_+}{\partial x}.  \label{eq:jdiff1}
\end{eqnarray}

We assume that the above equations, derived for a 2D lattice are also valid in a 3D hexagonal ice I$_h$ lattice. The hexagonal lattice is isotropic with a tetrahedral symmetry~\cite{RN10877, RN10878} when the small anisotropy of the ice lattice is ignored (around 1\% in the z direction~\cite{RN10943}). Before we proceed, we will first summarise the properties of the ice lattice that are related to this tetrahedral symmetry. A fundamental quantity of the lattice is $r_{oo}$, the distance between two oxygen atoms of neighbouring hydrogen bonded water molecules. In tetrahedral lattice, the density of ice $n_o$ relates to $r_{oo}$ as~\cite{RN10878}
\begin{equation}
n_o = \frac{ 3 \sqrt{3} }{8r_{oo}^3}. \label{eq:no}
\end{equation}
The mean distance $dx$ between the successive planes of water molecules is equal to
\begin{equation}
dx = \frac{ 2 r_{oo}}{\sqrt{3}}, \label{eq:tetra2}
\end{equation}
This equation is obtained from $n_o = 1/dx^3$ assuming an isotropic specimen. In \cite{jdp1} it is proven that 
\begin{equation}
p_{o,\parallel} = \frac{ e_{DL} r_{oo}}{\sqrt{3}}.  \label{eq:tetra3}
\end{equation}

The only unknown parameter is $v_+$. If $\tau_{+} $ is the average time interval of the ionic jumps, the average speed of these jumps in the x direction is 
\begin{equation}
v_+  = \frac{dx}{3\tau_{+}} 
\end{equation}   
taking into account that there are also jumps in the $y$ and $z$ direction. This velocity can now easily be related to the ionic diffusion constant ($D_+ = dx^2/(6\tau_{+}$)), leading to the expression
\begin{equation}
v_+  = \frac{2D}{dx}  \label{eq:tetra3}
\end{equation}   
Also the Einstein relation is relevant for the calculations. It relates the ionic diffusion coefficient D$_+$ to the ionic mobility , 
\begin{equation}
D_+ = \frac{k T\mu_{+}}{e_{\pm}} .  \label{eq:D+}
\end{equation}
with $k$ the Boltzmann constant and T is the absolute temperature.

Combining equations~\ref{eq:jdiff1} to ~\ref{eq:D+} into the equation of the flux $j_{+}$ (Eq.~\ref{eq:fluxD1}) leads to
\begin{equation}
j_{+} = \frac{\sigma_{+} }{e_{\pm}^2} (e_{\pm} E +  \frac{\Phi}{e_{DL}} P_s) - D_+ \frac{\partial n_+}{\partial x}  ,  \label{eq:j+}
\end{equation}
with the well-known $\Phi$ factor~\cite{RN10877} equal to
\begin{equation}
\Phi =   \frac{8 r_{oo} kT}{\sqrt{3}}. \label{eq:fluxD5} 
\end{equation}  

The first two terms are the classical terms found by Jaccard describing the flux density when no concentration gradients are present~\cite{jdp1, RN10943}. The third term is a classical diffusion term also used by Rhyzkin and Petrenko to describe screening effects in ice~\cite{RN1107,RN11870}. 

A similar equation can be written down for the negative ion, OH$^-$. Using $\sigma_-$ and $D_-$ as the conductivity and the diffusion constant of the negative ions respectively,  we obtain
\begin{equation}
j_{-} = \frac{\sigma_{-} }{e_{\pm}^2} (-e_{\pm} E -  \frac{\Phi}{e_{DL}} P_s) - D_-  \frac{\partial n_-}{\partial x}. \label{eq:j-}
\end{equation}

 \begin{figure}
\centering{\includegraphics[width=110mm]{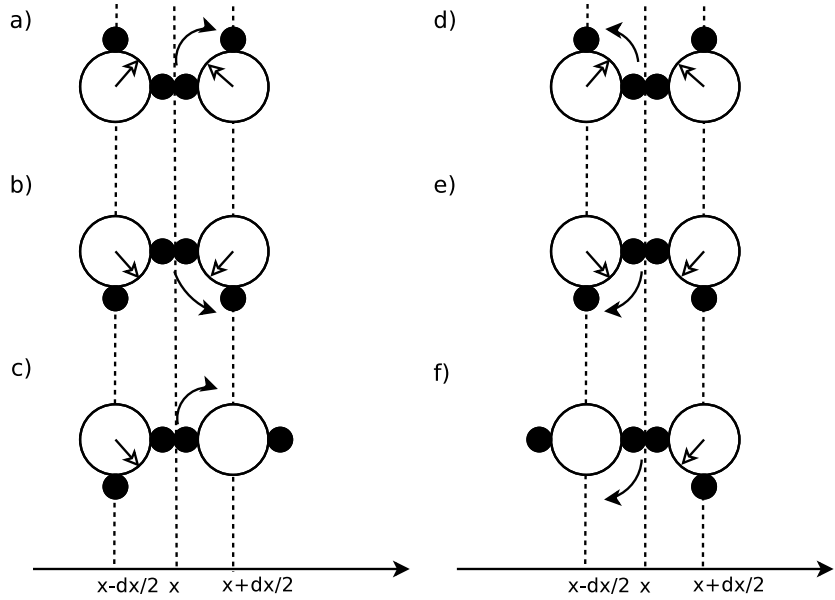}}
\caption{This 2D model shows the different ways a D defect can move in the positive x direction (a-c) and in the negative x direction (d-f). Notice that in case (c) and (f) only one H atom changes position during the turning of the water molecule.  } 
\label{fig:2Dmodel4} 
\end{figure}    

When an external field is applied, the flux density of the D defects will also be different in the positive and negative directions  (see Fig.~\ref{fig:2Dmodel4}). To find the correct equations it is important to define the particle density of the D defect, $n_D(x)$ more clearly. Because the horizontal D defects are divided over two water molecules we assign the defect to the water molecule that is left from the defect. This way, we avoid double counting of the horizontal defects. 
%Also, $n_D(x-dx/2)$ will contain both the vertical D defects as the ones situated at the right side of the position, the latter ones are drawn in Fig.~\ref{fig:2Dmodel4}(a-c). Opm.: de vertikale D defecten zullen spelen ook mee als de rechtermolecules goed geörienteerd zijn. 
The left watermolecule rule also makes clear when a D defect jumps from $x-dx/2$ to $x+dx/2$ and vice versa. In Fig.~\ref{fig:2Dmodel4}, cases a, b, d, e show the more classical jumps like in Fig.~\ref{fig:2Dmodel2}(b). However, cases c en f show also possible jumps, changing the actual position of the D defect. The similarity between Fig.~\ref{fig:2Dmodel4} and Fig.~\ref{fig:2Dmodel3} proves that the derivation of the D-defect and L-defect equations will use the same reasonings as for the ions, resulting in
\begin{eqnarray}
j_{D} = \frac{\sigma_{D} }{e_{DL}^2} (e_{DL} E -  \frac{\Phi}{e_{DL}} P_s) - D_D \frac{\partial n_D}{\partial x}, \label{eq:jD}\\
j_{L} = \frac{\sigma_{L} }{e_{DL}^2} (-e_{DL} E +  \frac{\Phi}{e_{DL}} P_s) - D_L \frac{\partial n_L}{\partial x},\label{eq:jL}
\end{eqnarray}
with $D_D$, $D_L$ the diffusion coefficients and $\sigma_{D}$, $\sigma_{L}$ the conductivities of the DL defects.

 \section{A one-dimensional homogeneous ice specimen in an external electric field}

The general solutions for a homogeneous ice specimen in an external field can be found in~\cite{jdp1}. A more practical approach is presented in this section by taking into account some special feature of most ice samples. 

Our first focus is the derivation of the polarisation density of the lattice structure $P_s$ when an external electric field $E$ is applied in the $x$ direction. The polarisation density $P_s$ is build up through the movement of bound charges. If $J_b$ is the current density of the bound charges, we get~\cite{RN13018}  
\begin{equation}
\frac{\partial P_s}{\partial t} = J_b = e_{DL} (j_{D}-j_{L}-j_{+}+j_{-}).   \label{eq:PlJ}
\end{equation} 
So, polarisation of an ice specimen is only possible if bound charges are moving in the specimen. 

In ice $I_h$ the ionic defect concentrations are far below the DL defects concentrations ($n_+, n_- \ll  n_D, n_L$) and this over a wide temperature range~\cite{RN13110}. This implies that the DL currents are dominating the polarisation density. But even more, experimental results show that the D defects are not mobile in ice because they are trapped near interstitials~\cite{RN10913}. This means that $D_D \ll D_L$ and $\sigma_D \ll \sigma_L$, leading to  $j_{D} \ll j_L$ and
 \begin{equation}
\frac{\partial P_s}{\partial t} = J_b \approx - e_{DL} j_{L}.   \label{eq:PlJ}
\end{equation} 

In DC, a steady state is reached,  so $J_b = 0$ and $j_L \approx 0$. In a homogeneous ice specimen, Eq.~\ref{eq:jL} reduces to 
\begin{equation} 
 j_{L} =  \frac{\sigma_{L} }{e_{DL}^2} (-e_{DL} E +  \frac{\Phi}{e_{DL}} P_s) \approx 0 . \label{eq:jL2} 
\end{equation} 
The electric current of L defects induced by the external electric field is in a first approximation compensated by an orientational diffusion current induced by polarisation of the ice specimen. 

The DC susceptibility of the solid ice lattice $\chi^{i}_s$ is defined  as  
\begin{equation}
P_{s} =  \epsilon_o \chi^{i}_s E.  \label{eq:PlNew} 
\end{equation}
Eq.~\ref{eq:jL2} describes this DC situation, leading to
\begin{equation}
\chi_s^i = \frac{e_{DL}^2}{\epsilon_o \Phi}, \label{eq:chili2}
\end{equation} 
a positive susceptibility which describes the experimental results accurately~\cite{RN10943}. 

Besides the polarisation, a homogeneous ice specimen is conducting a small electric current. The electrical current density is the sum of the current density of the bound and the free charges~\cite{RN13018}
\begin{equation}
\mathbf{ J} =  \mathbf{ J_b} +  \mathbf{ J_f} = e_{DL}(-\mathbf{j_{+}}+\mathbf{j_{-}} -\mathbf{j_L}) + e(\mathbf{j_{+}}-\mathbf{j_{-}}).  \label{eq:J}
\end{equation}

Because in steady state $J_b = 0$, we can derive the total current from Eqs.~\ref{eq:j+}, \ref{eq:j-} and \ref{eq:chili2}
\begin{equation}
\mathbf{ J} =  e(\mathbf{j_{+}}-\mathbf{j_{-}}) = \frac{e(\sigma_{+}+ \sigma_{-})}{e_{\pm}}  (1 + \frac{e_{DL}}{e_{\pm}}) E = \frac{(\sigma_{+}+ \sigma_{-})e^2}{e_{\pm}^2} E .  \label{eq:J}
\end{equation}
Also here we have a good match with the experimental results~\cite{RN10943}. The DC conductivity is determined by the smallest values of conductivity in the network. Both type of defects need each other to avoid netto polarisation of the ice specimen. Notice that although $ j_{L} \approx 0$ the L defects are still contributing to the total current density. Indeed, the netto current consists of three physical contributions
\begin{itemize}
\item an ohmic contribution of the ions $: (\sigma_{+}+ \sigma_{-}) E$,
\item a diffusional contribution of the ions because the ice is polarised by the L defects: $ \frac{e_{DL}(\sigma_{+}+ \sigma_{-})}{e_{\pm}}  E$,
\item a netto contribution of the L defects compensating the building up of extra polarisation by the ions $  \frac{e e_{DL}(\sigma_{+}+ \sigma_{-})}{e_{\pm}^2}  E$ 
\end{itemize}

\section{Conclusions}
Jaccards' equations for the defects in ice are derived using a new approach. It is based on an electrodiffusion model for the defects' kinetics. The polarisation of the ice specimen induces a directional diffusive current because it favours defects' orientations that diffuse opposite to the movement induced by the external electric field.  This approach has several didactical advantages compared to the approach in classical textbooks. It leads to both the ice equations and its standard solutions in a straightforward way and it makes a fundamental distinction between bound and free charges something the classical derivations failed to do. 

\section*{Acknowledgements} 
The author thanks Prof. Rui Qiao for pointing out some inconsistencies in my original work~\cite{jdp1}, which resulted in this paper.

\section*{References}

\bibliographystyle{unsrt}
\bibliography{literatuur}

\begin{thebibliography}{10}

\bibitem{RN1111}
J.~Sandblom.
\newblock Anomalous reactances in electrodiffusion systems.
\newblock {\em Biophysical journal}, 12(9):1118--1131, 1972.

\bibitem{RN10943}
V.~F. Petrenko and R.~W. Withworth.
\newblock {\em The physics of ice}.
\newblock Oxford University Press, 1999.

\bibitem{RN13131}
C.~Jaccard.
\newblock Mechanism of electrical conductivity in ice.
\newblock {\em Annals of the New York Academy of Sciences}, 125(A2):390--400,
  1965.

\bibitem{RN10877}
M.~Hubmann.
\newblock Polarization processes in the ice lattice. 1.approach by
  thermodynamics of irreversible processes - new experimental-verification by
  means of a universal relation.
\newblock {\em Zeitschrift Fur Physik B-Condensed Matter}, 32(2):127--139,
  1978.

\bibitem{RN1107}
M.~I. Ryzhkin, I.~A. Ryzhkin, and A.~V. Klyuev.
\newblock Screening of an electric field in water.
\newblock {\em JETP Letters}, 110(2):127--132, 2019.

\bibitem{jdp1}
John De~Poorter.
\newblock An improved formulation of jaccard’s theory of the electric
  properties of ice.
\newblock {\em The European Physical Journal B}, 92(7):157, 2019.

\bibitem{RN11870}
V.~F. Petrenko and I.~A. Ryzhkin.
\newblock Dielectric-properties of ice in the presence of space-charge.
\newblock {\em Physica Status Solidi B-Basic Research}, 121(1):421--427, 1984.

\bibitem{RN1110}
C.~Jaccard.
\newblock Thermoelectric effects in ice crystals.
\newblock {\em Physik der kondensierten Materie}, 1(2):143--151, 1963.

\bibitem{RN1109}
J.~Latham and Basil~John Mason.
\newblock Electric charge transfer associated with temperature gradients in
  ice.
\newblock {\em Proceedings of the Royal Society of London. Series A.
  Mathematical and Physical Sciences}, 260(1303):523--536, 1961.

\bibitem{RN10913}
M.~de~Koning.
\newblock First-principles modeling of lattice defects: advancing our insight
  into the structure-properties relationship of ice.
\newblock {\em Scientific Modeling and Simulations}, 15(1-3):123--141, 2008.

\bibitem{RN10914}
M.~de~Koning and A.~Antonelli.
\newblock On the trapping of bjerrum defects in ice i-h: The case of the
  molecular vacancy.
\newblock {\em Journal of Physical Chemistry B}, 111(43):12537--12542, 2007.

\bibitem{RN8152}
V.~G. Artemov and A.~A. Volkov.
\newblock Water and ice dielectric spectra scaling at 0 degrees c.
\newblock {\em Ferroelectrics}, 466(1):158--165, 2014.

\bibitem{RN11906}
A.~V. Klyuev, I.~A. Ryzhkin, and M.~I. Ryzhkin.
\newblock Generalized dielectric permittivity of ice.
\newblock {\em JETP Letters}, 100(9):604--608, 2015.

\bibitem{RN13018}
David~J. Griffiths.
\newblock {\em Introduction to Electrodynamics}.
\newblock Cambridge University Press, 2017.

\bibitem{RN11914}
S.~Cukierman.
\newblock Et tu, grotthuss! and other unfinished stories.
\newblock {\em Biochimica Et Biophysica Acta-Bioenergetics}, 1757(8):876--885,
  2006.

\bibitem{RN10878}
M.~Hubmann.
\newblock Polarization processes in the ice lattice. 2.approach by kirkwood
  theory - comparison with the results from thermodynamics of irreversible
  processes.
\newblock {\em Zeitschrift Fur Physik B-Condensed Matter}, 32(2):141--146,
  1978.

\bibitem{RN13110}
G.~P. Johari and E.~Whalley.
\newblock The dielectric-properties of ice ih in the range 272-133k.
\newblock {\em Journal of Chemical Physics}, 75(3):1333--1340, 1981.

\end{thebibliography}

\end{document}